\documentclass{aa}
\usepackage{graphics,graphicx}

\begin{document}

\title{Simultaneous single-pulse observations of radio pulsars}
\subtitle{III. The behaviour of circular polarization}
\author{A. Karastergiou 
	\inst{1,2}
	\and
	S. Johnston
	\inst{2}
	\and
	M. Kramer
	\inst{3}}

\institute{Max-Planck Institut f\"ur Radioastronomie, Auf dem H\"ugel
	69, 53121 Bonn, Germany \and School of Physics, University of
	Sydney, NSW 2006, Australia \and Jodrell Bank Observatory,
	University of Manchester, Macclesfield, Chesire SK11 9DL, UK }

\abstract{We investigate circular polarization in pulsar radio
emission through simultaneous observations of PSR B1133+16 at two
frequencies. In particular, we investigate the association of the
handedness of circular polarization with the orthogonal polarization
mode phenomenon at two different frequencies. We find the association
to be significant across the pulse for PSR B1133+16, making a strong
case for orthogonal polarization modes determining the observed
circular polarization. The association however is not perfect and
decreases with frequency. Based on these results and assuming emission
occurs in superposed orthogonal polarization modes, we present a
technique of mode decomposition based on single pulses. Average
profiles of the polarization of each mode can then be computed by
adding the individual mode-separated single pulses. We show that
decomposing single pulses produces different average profiles for the
orthogonal polarization modes from decomposing average
profiles. Finally, we show sample single pulses and discuss the
implications of the frequency dependence of the correlation of the
circular polarization with the orthogonal polarization mode
phenomenon.  
\keywords{pulsars: PSR B1133+16 -- polarization} }

\maketitle

\section{Introduction}\label{intro}

Perhaps one of the most elusive qualities of the highly polarized
pulsar radio emission is its circular polarization. Although it is
generally lower than its linearly polarized counterpart, it is amongst
the highest observed in astrophysical objects. Circular polarization
in pulsars has been attributed in the past to propagation effects on
the radiation as it travels through the pulsar magnetosphere,
geometrical effects or the emission mechanism itself.

Cordes et al. (1978) \nocite{crb78} were the first to point out an
association between the position angle (PA) of the linear polarization
and the handedness of the circular polarization. More specifically,
$90^o$ jumps in the PA were observed to occur simultaneously with
sense reversals in the circular polarization ($V$) in PSR
B2020+28. The authors argued that this could be adequately explained
by disjoint or superposed orthogonal polarization modes (OPMs). In the
disjoint case, each of the OPMs is emitted independently and only
individual modes are observed at a time. On the other hand, the
superposed case supports the idea that the observed radiation at any
instant is a mix of both modes, which are effectively emitted at the
same time, despite each mode having its own properties.  A number of
authors argued for the identification of OPMs as the natural
polarization modes in the pulsar magnetosphere (e.g. Barnard \& Arons
1986\nocite{ba86}, McKinnon 1997\nocite{mck97}, von Hoensbroech et
al. 1998\nocite{hlk98}, Petrova 2001\nocite{pet01}). Elliptically
polarized natural modes could then provide the origin for the circular
polarization as a propagation effect.

Stinebring et al. (1984a, b)\nocite{scr+84} studied the polarization
behaviour of single pulses from a number of pulsars using grey-scale
density plots to investigate the fluctuations of the PA, the linear
polarization, $L$, and $V$. They noticed that PA jumps often occur at
the same pulse longitude at 430 and 1404 MHz and that the orthogonal
polarization mode (OPM) phenomenon was becoming more prevalent at
higher frequencies. They also noticed that at the phase bins where OPM
jumps occur, the degree of linear polarization was significantly
reduced, which led to the idea that, at any given instant, the
radiation observed is the incoherent sum of OPMs. This has the
advantage of naturally reducing the linear polarization and
attributing its fluctuations to fluctuations in the polarization of
the OPMs, which have a constant degree of linear polarization. In
contrast, if OPMs were emitted independently, the emission mechanism
would have to intrinsically produce a largely fluctuating degree of
linear polarization. McKinnon \& Stinebring (1998\nocite{ms98})
realized that the assumption of incoherent addition of OPMs can be
used to decompose an integrated pulse profile into the profiles of
each OPM. By accounting for the effects of instrumental noise, this
method can be used to reproduce the histograms of the distributions of
the PA and $L$ from Stinebring et al. (1984b). Mode decomposed average
profiles of PSRs B0525+21 and B2020+28 were shown in McKinnon \&
Stinebring (2000\nocite{ms00}) using their technique on the average
profiles of these pulsars.

In this paper, the association of circular polarization with the
orthogonal polarization modes is examined in PSR B1133+16 at 1.41 GHz
and 4.85 GHz, having studied the correlation of OPMs between these two
frequencies in Karastergiou et al. (2002, hereafter Paper
II)\nocite{kkj+02}. We make use of contingency tables to quantify this
association at each frequency, which we compare to the findings of
Cordes et al. (1978). We develop a method to decompose single pulses
into the individual OPMs. The method is simple and only assumes that
the observed Stokes $I$, $Q$, $U$, $V$ are the incoherent sum of the
100\% polarized OPMs. We show the different mode-decomposed profiles
of PSR B1133+16 by following our method and that of McKinnon \&
Stinebring (2000) and discuss the differences. Finally, we present
example single-pulses from our simultaneous data set to highlight the
observed effects and we discuss possible interpretations.

\section{The data}

We use the same dataset on PSR B1133+16 as in Paper II. It consists of
4778 pulses at 4.85 and 1.41 GHz observed with the Effelsberg and
Jodrell Bank radio telescopes respectively. Full details of the
observing, calibration and data reduction can be found in Paper
II. The mean S/N of $V$ detected in the single pulses is
$<<1$. However, in a significant number of cases, the signal in $V$
exceeds the {\it rms} of the noise by many times at both
frequencies. At 4.85 GHz, where the flux density of this pulsar is
much less than 1.41 GHz, the sensitivity of the Effelsberg telescope
largely compensates. This is evident from the S/N of $V$ shown in
Fig. 1 and favours our results to be intrinsic to the pulsar emission
rather than products of the instrumental noise.

\subsection{Distributions of $V$ and $V/I$}
\begin{figure}[t]
\centerline{
\resizebox{\hsize}{!}{\includegraphics{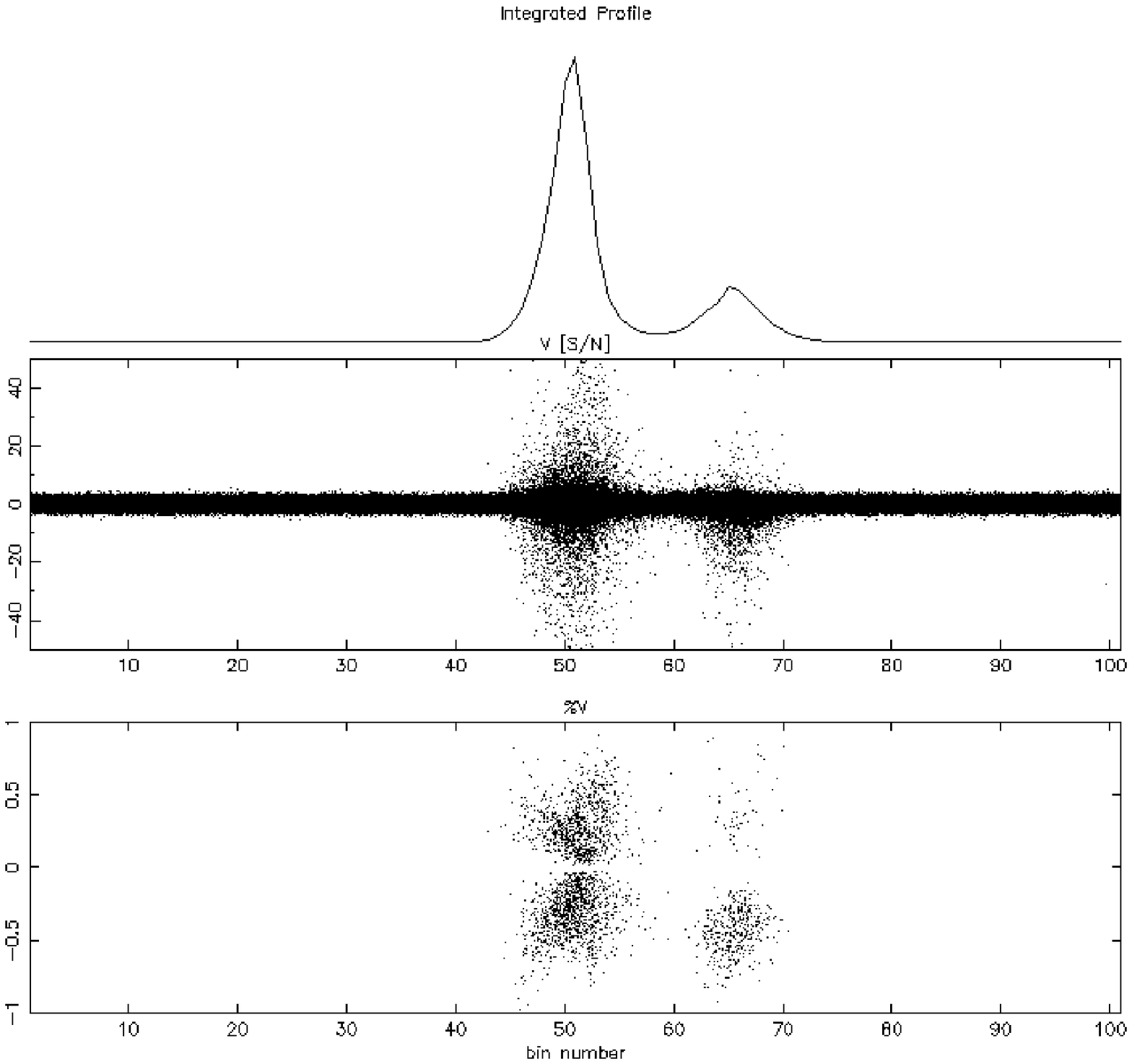}}
}
\centerline{
\resizebox{\hsize}{!}{\includegraphics{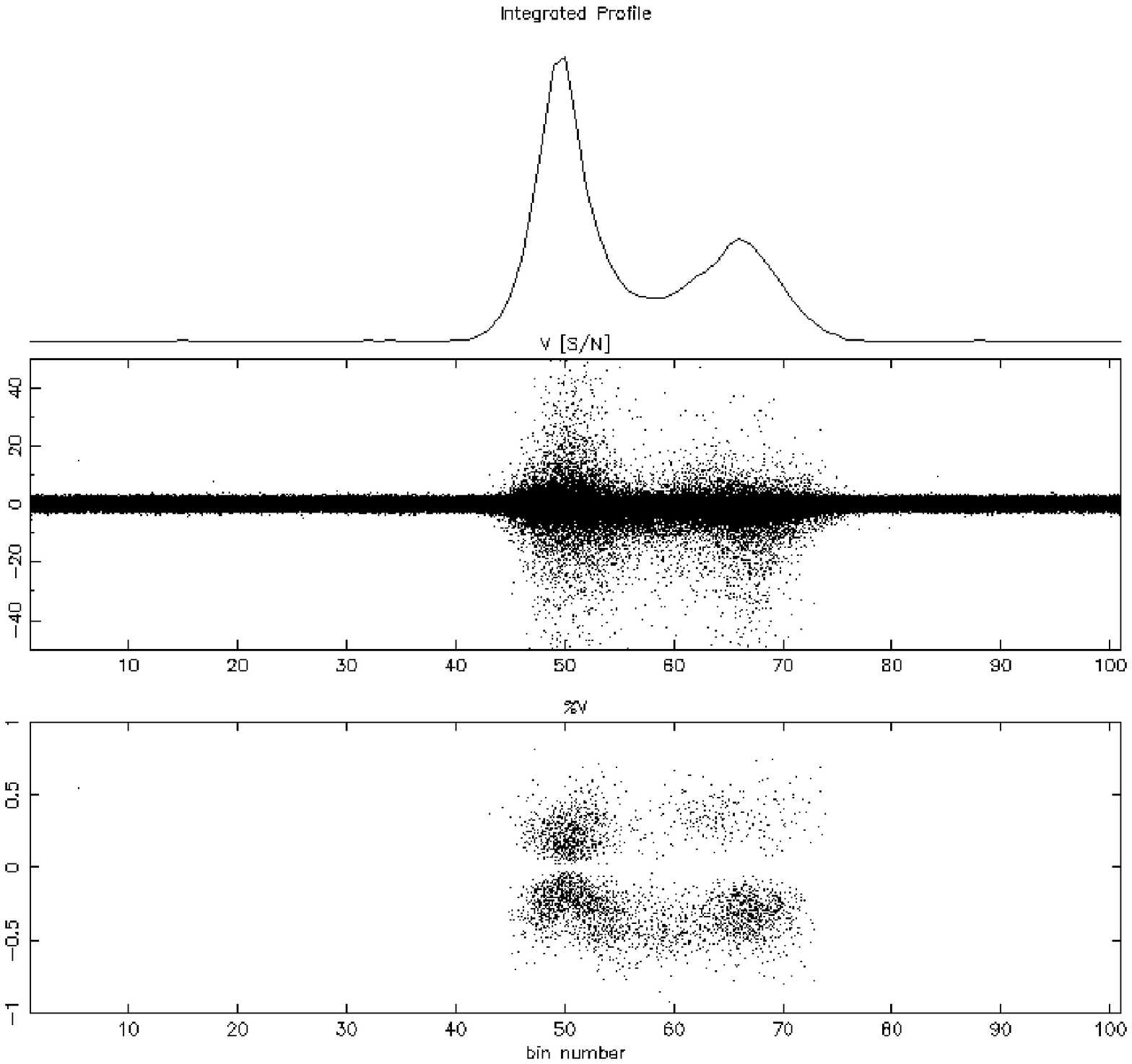}}
}
\caption[$V$ and $V/I$ distributions in PSR B1133+16] {The
distribution of $V$ and $V/I$ in PSR B1133+16, at 4.85 GHz [top] and
1.41 GHz [bottom]. Each plot has three horizontal panels. The top panel
shows the total power ($I$) integrated pulse profile, the middle panel
shows the distribution of flux densities of $V$ in units of the {\it
rms} and the bottom panel shows the distribution of the fraction
$V/I$. In the bottom panel, a threshold has been applied to exclude
the noise: only cases where the flux density of $V$ is more than
$5\times rms$ are shown. }
\label{vdist}
\end{figure}
\begin{figure}[t]
\resizebox{\hsize}{!}{\includegraphics[angle=-90]{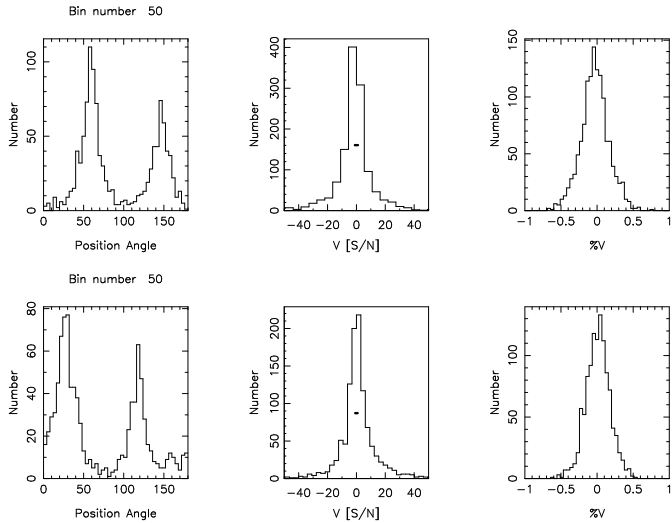}}
\caption[PA, $V$ and $V/I$ distributions in PSR B1133+16(bin 50)]
{The distribution of PA, $V$ and $V/I$ in bin 50, near the peak of the
leading component, at 4.85 GHz [top] and 1.41 GHz [bottom].}
\label{v50}
\end{figure}
\begin{figure}[t]
\resizebox{\hsize}{!}{\includegraphics[angle=-90]{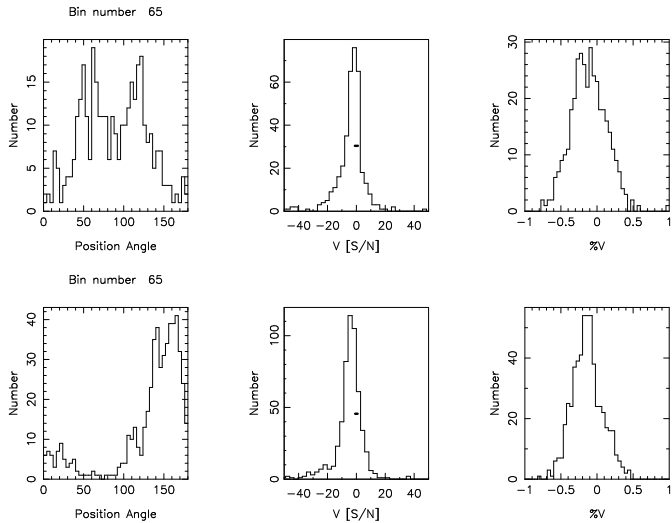}}
\caption[PA, $V$ and $V/I$ distributions in PSR B1133+16(bin 50)] {The
distribution of PA, $V$ and $V/I$ in bin 65, near the peak of the
trailing component, at 4.85 GHz [top] and 1.41 GHz [bottom].}
\label{v65}
\end{figure}

To investigate the single-pulse behaviour of circular polarization,
the value of $V$ is plotted against the corresponding pulse phase for
every single pulse. Such density plots reveal the pulse to pulse
fluctuations of $V$. Fig. \ref{vdist} shows two such plots for the
single-pulse data of PSR B1133+16 observed simultaneously at 4.85 GHz
and 1.41 GHz. The density plots reveal that the distributions of $V$
at each pulse phase bin range over a few tens of {\it rms} units at both
frequencies. 


In the leading component, it has been shown that the OPMs are almost
equally strong and dominate equally often at both frequencies (Paper
II). The distribution of $V$ in this component shows an almost equally
dense population of positive and negative values. On the other hand,
the trailing component is almost always dominated by one OPM and the
respective $V$ distribution favours negative values. Under incoherent
addition of OPMs, the observed $V$ is the difference between the $V$
of each OPM and therefore the above observation is in accordance with
a particular handedness of $V$ being associated with a certain OPM.

The seemingly bimodal distributions of $V/I$ in the leading component
in Fig. \ref{vdist} are caused by the threshold applied to exclude the
noise. Fig. \ref{v50} shows the specific distributions of $V$ and
$V/I$ together with the PA histogram of bin 50. The bottom row
corresponds to the 1.41 GHz data and the top row to the 4.85 GHz
data. The threshold on $V$ is not applied here, in order to
demonstrate how much of the distribution lies within the instrumental
noise, the $rms$ of which is represented by a small horizontal bar
inside the $V$ distribution. Only the pulses in which the
signal-to-noise (S/N) of the total power exceeds 2 {\bf and} the
fractional linear polarization exceeds 20\% are considered, in order
to exclude ambiguous PA values. Bin 50 exhibits a clear bimodal PA
distribution at both frequencies and at the same time a mean $V$ value
very close to 0. On the other hand, Fig. \ref{v65} shows bin 65 from
the trailing pulse component, that presents a slightly different
picture. The PA distributions at both frequencies are dominated by one
OPM (at 4.85 GHz this is not entirely clear due to the small
population of points that survive the threshold) and the $V$
distributions have a significantly non-zero negative mean. This
picture is consistent with incoherently superposed OPM, each mode
favouring a particular handedness of $V$: in bin 50 where both modes
dominate equally often, $V$ is distributed around 0, which is not the
case for bin 65, where one mode essentially dominates and the
handedness of $V$ associated with this mode gives mean $V$ its
negative value.

Figs. \ref{v50} and \ref{v65} show a difference between the
distributions in $V$ and $V/I$. All four $V/I$ histograms can be
adequately approximated by Gaussian distributions. In bin 50, the best
fit has a mean of $-2.5\%$ and a $\sigma$ of $19\%$ at 4.85 GHz and a
mean of $0.6\%$ and a $\sigma$ of $18\%$ at 1.41 GHz. In bin 65 on the
other hand, the best fit has a mean of $-12\%$ and a $\sigma$ of
$27\%$ at 4.85 GHz and a mean of $-13\%$ and a $\sigma$ of $24\%$ at
1.41 GHz. These broad distributions are consistent with those seen in
other pulsars (Stinebring et al. 1984a, b\nocite{scw+84}) and are
largely dominated by instrumental noise (McKinnon 2002).

In summary, the density plots of the circular polarization in
Fig. \ref{vdist} and the individual distributions in Fig. \ref{v50}
and Fig. \ref{v65}, together with the OPM analysis presented in Paper
II, provide evidence that each OPM is associated with a certain
handedness of $V$. This association, however, is obviously not
perfect, and we discuss this in the following.

\subsection{The $V$ - OPM correlation}

In Cordes et al. (1978), observations of PSR B2020+28 at 404 MHz were
used to investigate a possible connection between the handedness of
$V$ and the PA value. To this goal, they plotted $V$ versus PA in the
single pulses they observed. In their paper, they give an example of a
particular pulse longitude where the distribution of PAs is bimodal,
in other words a phase bin where the dominant OPM alternates from
pulse to pulse. The plot of $V$ versus PA for this bin demonstrates a
clear correlation between the handedness of $V$ and the OPM.

In Paper II, a detailed study of the frequency dependence of the OPMs
in PSR B1133+16 demonstrated that, at higher frequencies, the OPMs are
seen to dominate almost equally often and the individual fluxes of the
OPMs become more equal to each other. Given this frequency evolution,
we investigate whether the $V$ - OPM association is also frequency
dependent. Instead of $V$ - PA plots, we follow a different route,
namely to display the tips of the polarization vectors of all the
single pulses from specific bins on the Poincar$\acute{\rm e}$ sphere. The
coordinates of a given point on the surface of the sphere are given by
two angles, the latitude ($\tan^{-1}V/L$) and longitude ($2PA$). The
latitude is a measure of the ellipticity of the polarization state:
values of $0^o$ correspond to linearly polarized states whereas values
of $\pm 90^o$ correspond to purely circular states of opposite
handedness.
\begin{figure*}[t]
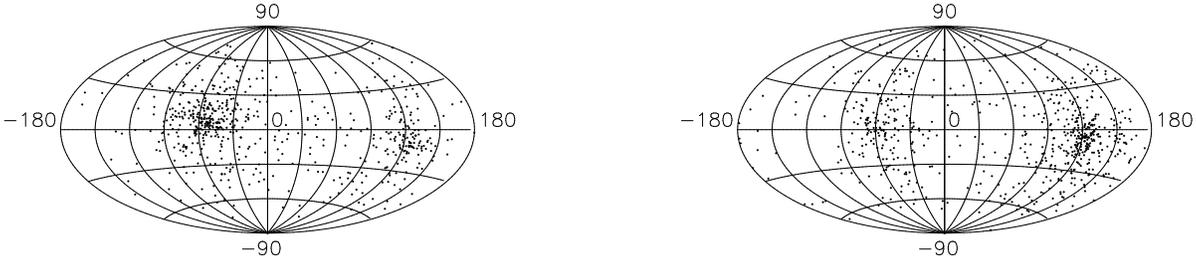

\begin{minipage}[b]{0.5\linewidth}
\includegraphics[angle=-90,width=6.8cm]{H4198F4a.ps}
\end{minipage}%
\begin{minipage}[b]{0.5\linewidth}
\includegraphics[angle=-90,width=6.8cm]{H4198F4b.ps}
\end{minipage}
\caption
{Hammer-Aitoff projection of the Poincar$\acute{\rm e}$ sphere, where
the polarization states of a phase bin 50 from the single pulses are
depicted as the tip of the polarization vector on the
Poincar$\acute{\rm e}$ sphere. The left panel corresponds to 1.41 GHz
and the right panel to 4.84 GHz. Bimodal PA distributions are evident
in both plots. }
\label{poin}
\end{figure*}
Fig. \ref{poin} shows the bimodal distributions of PA in phase bin 50,
the modes being $\sim90^o$ apart in PA. It also shows that the
majority of points in each mode prefer a certain handedness in $V$ and
therefore lie on either side of the equator ($V=0$). Comparing
Fig. \ref{poin} to Fig. 6 Cordes et al. (1978), reveals that the
association of $V$ with PA is not as clear in PSR B1133+16 at these
frequencies as it was in PSR B2020+28 at 404 MHz. A quantitative test
is therefore necessary.

In the scheme of OPM superposition, it is very important to stress
again that the observed PA is that of the dominant mode, regardless of
the flux density difference of the two modes. Therefore, by testing
the PA vs. the sign of $V$ association, one is effectively testing the
association of the handedness of $V$ with OPM. For each phase bin of
the pulse and for the total number of pulses, we quantize the observed
quantities according to the following $3\times 2$ possibilities:
dominant mode 1, dominant mode 2 or not defined PA, and positive-$V$
or negative-$V$. By composing such {\it contingency tables} for each
phase bin, it is possible to quantify the association between the
handedness of $V$ and the dominant OPM. An example for bin 50 at each
observed frequency is shown in Table 1.
\begin{table}
\begin{center}
\begin{tabular}{c|ccc}
4.85 GHz & Mode 1 & Mode 2 & Undefined \\ \hline
$V +$ & 390 & 452 & 1366 \\
$V -$ & 659 & 436 & 1475 \\
\end{tabular}
\\
\begin{tabular}{c|ccc}
1.41 GHz & Mode 1 & Mode 2 & Undefined \\ \hline
$V +$ & 269 & 632 & 1506 \\
$V -$ & 442 & 386 & 1563 \\
\end{tabular}
\end{center}
\caption{Contingency tables for bin 50 of the PSR B1133+16 data,
showing the frequency of occurrence of all possible combinations in the
single pulses. The top table corresponds to the 4.85 GHz and the bottom
table to the 1.41 GHz data.}
\end{table}
The numbers have been obtained by applying a two step process. Where
the S/N of $L$ in the given bin is less than 2.5, we are unable to
compute the PA and therefore classify these as ``Undefined'' in Table
1. Then, we compare the observed PA in each single pulse to the PA of
the integrated profile in this bin, to determine which mode is
dominating at this particular instant. From the numbers presented in
Table 1, it is possible to compare the observed cases to the null
hypothesis, by using the formula
\begin{equation}
D=\pm\frac{(Obs-Null)^2}{Null}
\end{equation}
where $Obs$ are the observed numbers and $Null$ the null
hypothesis. Negative values denote a deficit of the observed cases
from the null hypothesis. For every bin such as the one presented in
Table 1 there will be four values of $D$, one for each table
element.

\begin{figure}[t]
\resizebox{\hsize}{!}{\includegraphics[angle=-90]{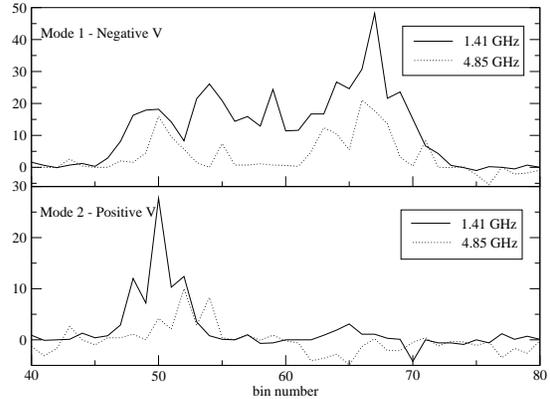}}
\caption{$D$ plotted against the pulse phase bin for the Mode 1
- $V-$ and Mode 2 - $V+$ cases. The solid line corresponds to 1.41 GHz
and the dotted line to 4.85 GHz. The negative $D$ values
signify a deficit from the null hypothesis.}
\label{Vcont}
\end{figure}

We perform this analysis across the whole profile of PSR
B1133+16. What we find is that there is an excess from the null
hypothesis of negative $V$ cases across the pulse for the most common
OPM (Mode 1) and a positive $V$ excess from the null hypothesis for
the other OPM (Mode 2). Fig. \ref{Vcont} shows $D$ as measured by
Eq. (1), plotted across the pulse. The top panel of the plot shows the
$D$ values for the association between Mode 1 and negative $V$,
whereas the bottom panel shows the same for Mode 2 and positive $V$,
at both frequencies. From the definition of $D$ we find that values
over 5 are statistically significant.  Two facts are evident from
these plots. First, the values of $D$ are generally large enough to
support a strong correlation between the handedness of $V$ and the
dominant OPM. Second, the $D$ values are consistently larger at 1.41
GHz than at 4.85 GHz, indicating that the correlation is weaker at
4.85 GHz. Also, as we have seen before, Mode 2 hardly ever dominates
in the trailing pulse component and therefore the corresponding $D$
values are all close to zero.

\section{OPM separation}

Our statistical analysis of the dual frequency PSR B1133+16 data has
shown us that there is a frequency dependent correlation between the
handedness of $V$ and the dominant OPM. The contingency tables,
however, show that there is a significant population of cases where
the handedness of $V$ is opposite from the expected. In other words,
the $V$ - OPM association is not perfect. This is a very important
result with serious implications on the technique of decomposing pulse
profiles into the individual OPMs. The significance lies in the fact
that each OPM that dominates in a given single pulse may be observed
instantaneously with either handedness of $V$. However, averaging the
single pulses and decomposing the average profile is based on an
assumption that is not met, namely that each OPM is clearly associated
with a particular sign of $V$ in {\bf all} the single pulses. To
demonstrate this, we detail the mathematics of mode-decomposition in
single pulses and show the difference between the obtained profiles by
decomposing and then averaging, and averaging and then decomposing the
PSR B1133+16 data.

The orthogonality of the modes means that their polarization states at
a given instant are represented by antiparallel vectors on the
Poincar$\acute{\rm e}$ sphere. By definition therefore the PAs can
only take values which are exactly $90^o$ apart. Incoherent
superposition allows the addition of the individual Stokes
parameters. The mathematical process is as follows:\\ The observed
Stokes parameters can be written in terms of the OPMs
\begin{eqnarray}\label{inco}
I&=&I_1+I_2\nonumber\\
Q&=&Q_1+Q_2\nonumber\\
U&=&U_1+U_2\nonumber\\
V&=&V_1+V_2\nonumber\\
L&=&L_1+L_2\nonumber \ ,\\
\end{eqnarray}
where $L_i=\sqrt{Q_i^2+U_i^2}$. The 100\% polarized modes add up to
produce the observed, partially polarized radiation, where $L^2+V^2
\neq I^2$. Defining $P$ as the polarized intensity $\sqrt{L^2+V^2}$,
and considering that the antiparallel vectors must have opposite
senses of circular polarization, the orthogonal polarization states
imply that,
\begin{eqnarray}
\frac{L_i}{I_i}=d_L=\frac{L}{P}\ ,\\
\frac{|V_i|}{I_i}=d_V=\frac{V}{P}\ .
\end{eqnarray}
Also, the OPMs are 100\% polarized, which means that
\begin{equation}
d_L^2+d_V^2=1\ {\rm or}\ d_V=\sqrt{1-d_L^2}\ .
\end{equation}
Taking into account the antiparallel vectors of the OPM polarization
states and ignoring $Q$ and $U$ for the moment, Eq. (\ref{inco}) can be
written as
\begin{eqnarray}
I&=&I_1+I_2\nonumber\\
L&=&d_L(I_1-I_2)\nonumber\\
V&=&\sqrt{1-d_L^2}(I_1-I_2)\nonumber \ ,\\
\end{eqnarray}
which are sufficient to solve for $I_1$ and $I_2$, so that
\begin{eqnarray}\label{Isplit}
I_1&=&0.5\cdot(I+\frac{L}{d_L})\nonumber \\
I_2&=&0.5\cdot(I-\frac{L}{d_L})\nonumber \  .\\
\end{eqnarray}
and therefore, Eqs. (3) and (4) lead to
\begin{eqnarray}
L_1&=&d_L\cdot I_1\nonumber\\
L_2&=&d_L\cdot I_2\\ \label{vsplit}
V_1&=&d_V\cdot I_1\nonumber \\
V_2&=&-d_V\cdot I_2\ .\\ \label{lsplit}\nonumber
\end{eqnarray}
To decompose $Q$ and $U$, the PA is used. The PA of the
strongest mode should be equal to the observed PA, and the PA of the
weak mode should be orthogonal to this. Therefore,
\begin{equation}
\frac{U}{Q}=\frac{U_1}{Q_1}=\frac{-U_2}{-Q_2}=\frac{U_2}{Q_2}\ ,
\end{equation}
from which $Q_i$ and $U_i$ can be determined as
\begin{eqnarray}
Q_i&=&\frac{L_i}{L}\cdot Q\nonumber \\
U_i&=&\frac{L_i}{L}\cdot U\nonumber \ .\\
\end{eqnarray}

The above method shows that there is a unique way to decompose the
received Stokes parameters into the Stokes parameters of the
individual orthogonal modes. In practice, special care must be taken
in the cases of very small $L$ and $V$ to avoid significant
errors. The equations also express the intrinsic assumption of this
method that the modes are orthogonal, so the circular polarization
sense is clearly opposite in each mode in a given single
pulse. However, the sense of $V$ that is associated with a particular
OPM may change from pulse to pulse.

In Stinebring \& McKinnon (2000), the average profiles of PSRs B0525+21
and B2020+28 are decomposed into the OPMs, using the assumption of
superposed OPM emission. We apply our method of single-pulse
decomposition and their method of average profile decomposition to PSR
B1133+16 to identify and interpret the differences.
\begin{figure*}[t]
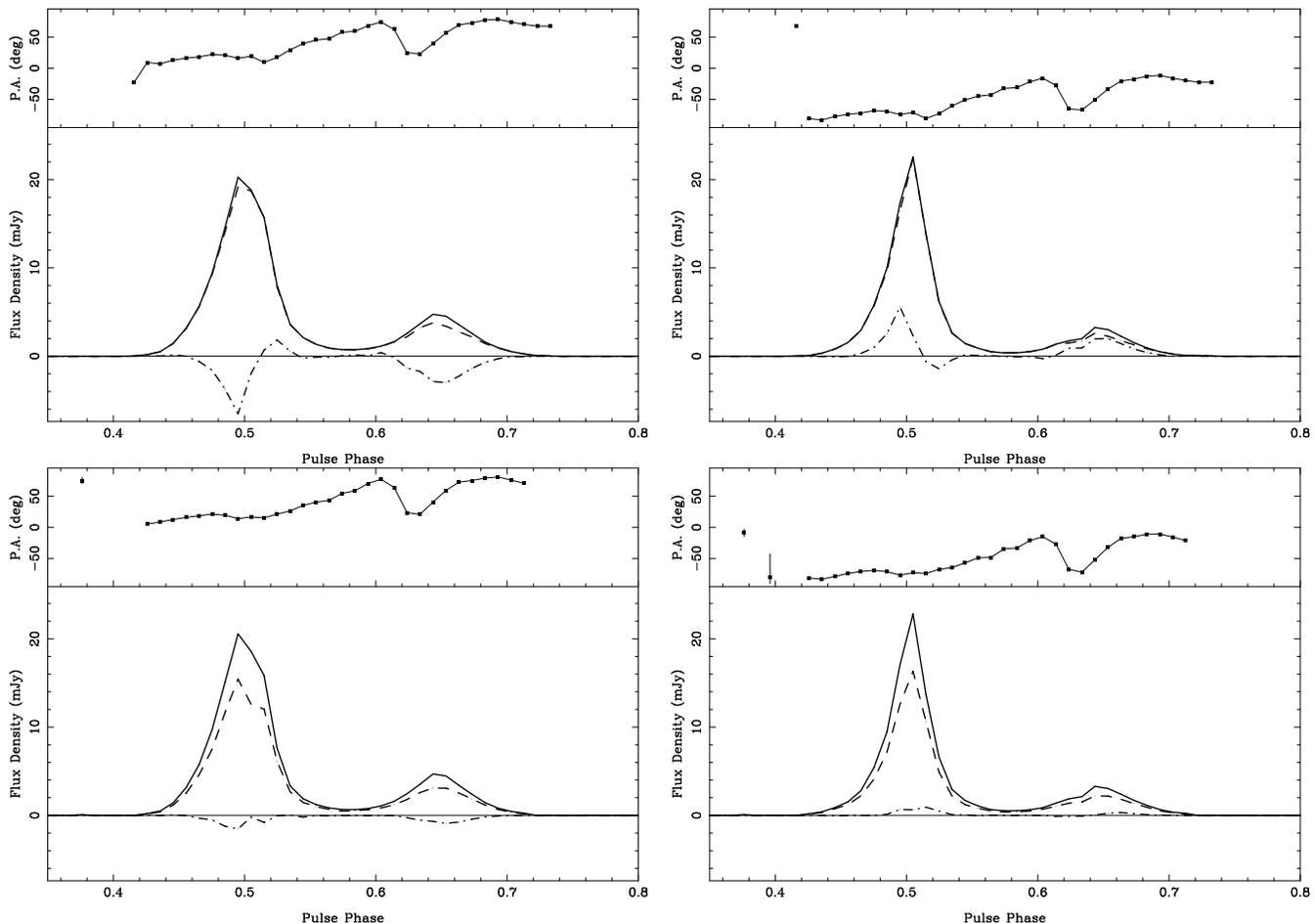

\begin{minipage}[b]{0.5\linewidth}
\includegraphics[angle=-90,width=8.7cm]{H4198F6a.ps}
\end{minipage}%
\begin{minipage}[b]{0.5\linewidth}
\includegraphics[angle=-90,width=8.7cm]{H4198F6b.ps}
\end{minipage}
\begin{minipage}[b]{0.5\linewidth}
\includegraphics[angle=-90,width=8.7cm]{H4198F6c.ps}
\end{minipage}%
\begin{minipage}[b]{0.5\linewidth}
\includegraphics[angle=-90,width=8.7cm]{H4198F6d.ps}
\end{minipage}
\caption{The top row of plots shows the average pulse profile of PSR
  1133+16 at 4.85 GHz, decomposed into the profiles for Mode 1 (left)
  and Mode 2 (right). The bottom row is the result of the
  decomposition into the individual OPMs of each single pulse, which
  are then averaged. In each plot, the main panel shows the total
  power $I$ (solid), the linear polarization $L$ (dashed) and the
  circular polarization $V$ (dashed-dotted). The secondary panel on
  top shows the PA.}
\label{f1d}
\end{figure*}
\begin{figure*}[t]
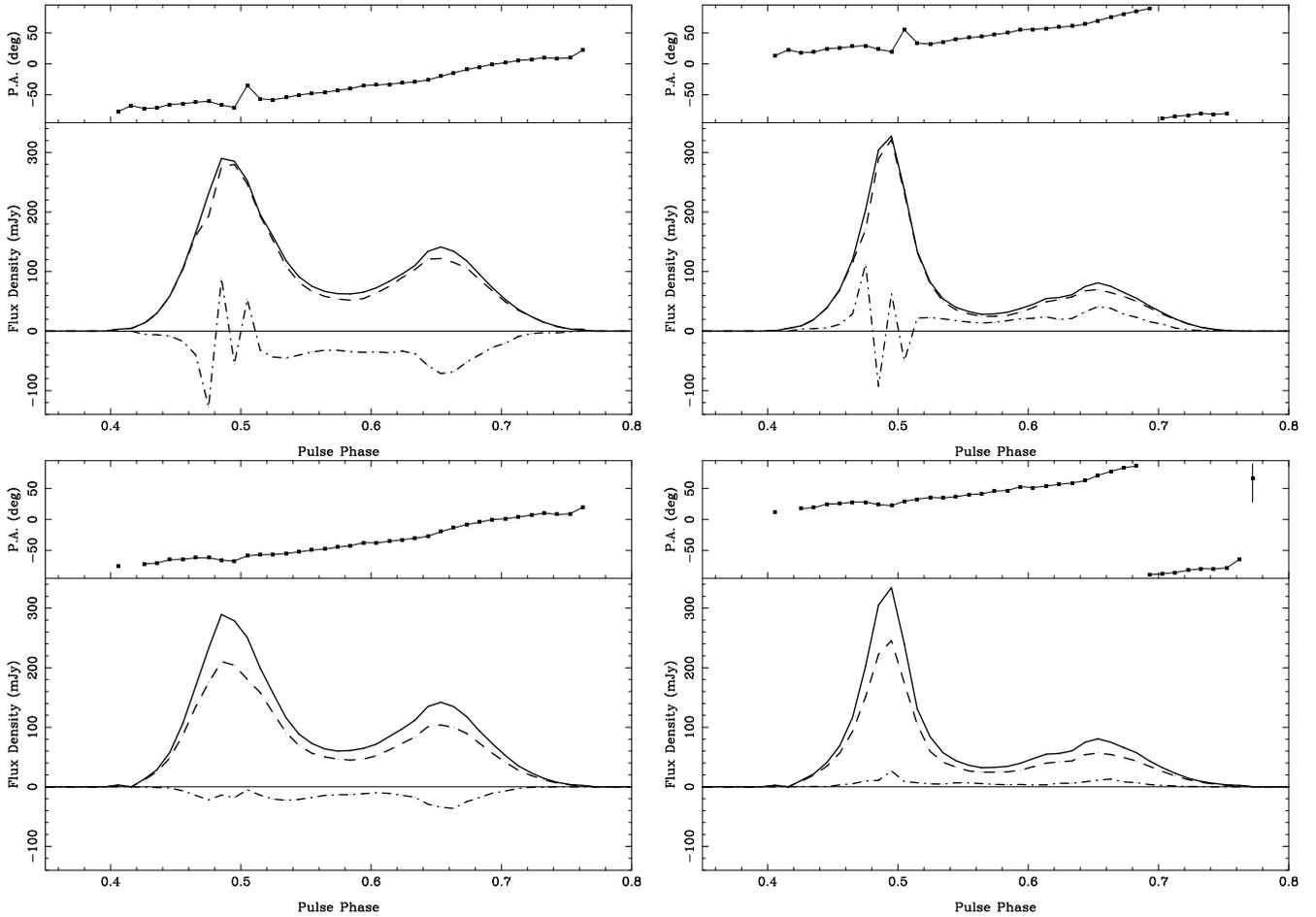

\begin{minipage}[b]{0.5\linewidth}
\includegraphics[angle=-90,width=8.7cm]{H4198F7a.ps}
\end{minipage}%
\begin{minipage}[b]{0.5\linewidth}
\includegraphics[angle=-90,width=8.7cm]{H4198F7b.ps}
\end{minipage}
\begin{minipage}[b]{0.5\linewidth}
\includegraphics[angle=-90,width=8.7cm]{H4198F7c.ps}
\end{minipage}%
\begin{minipage}[b]{0.5\linewidth}
\includegraphics[angle=-90,width=8.7cm]{H4198F7d.ps}
\end{minipage}
\caption{Same as Fig. \ref{f1d} for 1.41 GHz.}
\label{f2d}
\end{figure*}
Figs. \ref{f1d} and \ref{f2d} show the results of the two different
approaches to OPM-decomposition at 4.85 GHz and 1.41 GHz
respectively. First of all, the averaged profiles created using the
two methods are obviously different. The average circular polarization
in the case of the single-pulse decomposition has a constant
sense across the profile, which is not true for the average profile
decomposition. Also, it can be seen that the average flux density of
$V$ in each of the modes is considerably smaller in the case of
single-pulse decomposition. Another aspect of the observed differences
between the profiles is the linear polarization, which is $\sim100\%$
in the decomposed average profiles and significantly less in the
decomposed single-pulse profiles. We claim that the reduced degree of
polarization is due to the spread in the PA distributions. We have
found that a $45^o$ spread in the PA distribution of a particular
phase bin, results in a $\sim20\%$ decrease in the mean linear
polarization after averaging a few hundreds of single pulses. For this
reason, it is impossible to obtain mode-separated average profiles
constructed from decomposed single pulses that exhibit $100\%$
polarization unless the PA distributions are delta-functions.

\section{Simultaneous pulses}

The significant correlation between $V$ and OPM is in agreement with
the initial result for PSR B2020+28 (Cordes et al. 1978) and
constitutes evidence that circular polarization is closely linked to
the OPM phenomenon. However, we have demonstrated that decomposing
single pulses into the OPMs and averaging produces different
mode-separated profiles from averaging single pulses first and then
decomposing the resulting profile. This happens because the
correlation of $V$ with OPM, although significant, is not perfect, and
becomes worse with frequency. Obviously, the entire mechanism that
determines the polarization we observe is not well understood.

In Karastergiou et al. (2000\nocite{khk+01}, Paper I) and Paper II it
was shown that single pulses of PSRs B1133+16 and B0329+54 observed
simultaneously at two, widely spaced frequencies may often disagree in
terms of the dominant OPM. Despite these frequent cases of
disagreement, the cases of agreement were statistically more
significant and therefore revealed a high degree of correlation in
OPMs between the observed frequencies. The correlation of OPMs between
frequencies is facilitated by the fact that the PAs of the OPMs are,
by definition, orthogonal. Therefore, it is possible to distinguish
the prevailing OPM at both frequencies and count the cases of
agreement and disagreement. Unfortunately, the observed $V$ has wide
unimodal distributions around a usually very small mean. An
investigation of the correlation of the handedness of $V$ between the
frequencies, is therefore dominated by the mean and the $rms$ of the
$V$ distribution. As an example, the bins in the trailing pulse
component of PSR B1133+16 as seen in Fig. 1, show broad $V$
distributions of a clearly negative mean. A test of the association in
handedness of $V$ between the frequencies in this component obviously
shows a high correlation, forced by the underlying distributions
themselves. As for the leading pulse component, we have demonstrated
that each mode is clearly associated to a particular handedness in $V$
at a given frequency. The high degree of the correlation of the OPMs
between the frequencies therefore also results in a high correlation
in the handedness of $V$. To summarize, we see a high correlation in
the handedness of $V$ between the observed frequencies, which is what
we expect from our previous knowledge. This correlation, however, does
not add significantly to our interpretation of the observations.

\begin{figure}[t]
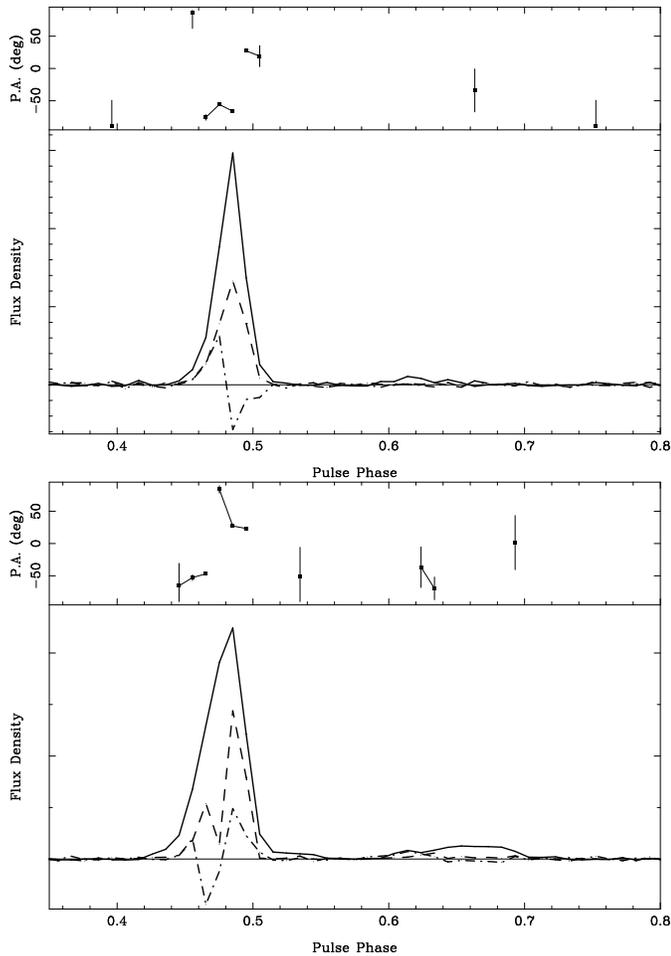

\centerline{
\resizebox{\hsize}{!}{\includegraphics[angle=-90]{H4198F8a.ps}}
}
\centerline{
\resizebox{\hsize}{!}{\includegraphics[angle=-90]{H4198F8b.ps}}
}
\caption[Single pulse 1562] {Single pulse number 1562 from PSR
B1133+16, at 4.85 GHz (top) and 1.41 GHz (bottom). The solid line
represents the total power, the dashed line the linear polarization,
and the dashed-dotted line the circular polarization. The PA jumps in
the same direction at both frequencies, but the circular polarization
follows a swing of opposite handedness between the frequencies.}
\label{p1562}
\end{figure}
\begin{figure}[t]
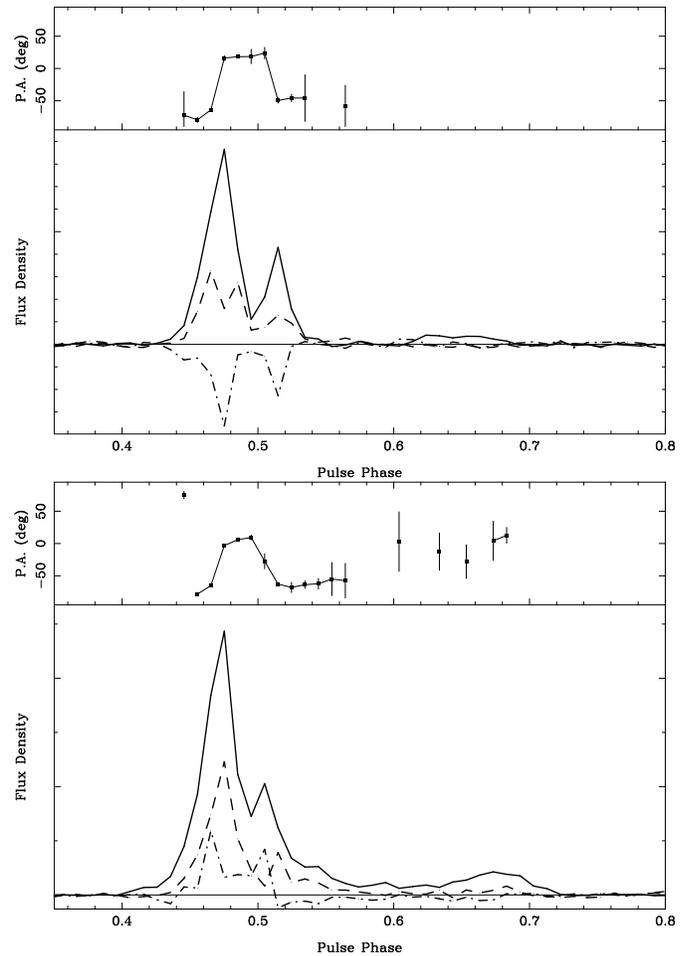

\centerline{
\resizebox{\hsize}{!}{\includegraphics[angle=-90]{H4198F9a.ps}}
}
\centerline{
\resizebox{\hsize}{!}{\includegraphics[angle=-90]{H4198F9b.ps}}
}
\caption[Single pulse 2438] {Single pulse number 2438 from PSR
B1133+16, at 4.85 GHz (top) and 1.41 GHz (bottom). The
circular polarization disagrees in handedness between the
frequencies. The $V$ flux densities match very well, showing a peak at
each phase bin of the PA jump.}
\label{p2438}
\end{figure}
Given the simultaneity of our data, we have the opportunity to inspect
single pulses as a whole observed at both frequencies. 
Two example single-pulses are provided in Figs. \ref{p1562} and
\ref{p2438}. Fig. \ref{p1562} shows pulse number 1562 from PSR
B1133+16 at 4.85 GHz (top panel) and 1.41 GHz (bottom panel). The
leading pulse component shows the characteristic swing and change of
handedness in $V$ usually associated with core components. The sense
of the swing is opposite between the two frequencies although the PAs
agree. They both show a PA jump in the same direction, although the
jump happens slightly earlier at the lower frequency. In pulse 2438 in
Fig. \ref{p2438}, an OPM jump occurs in bin 48 and the PA jumps back
again in bin 52, at both frequencies. Both jumps correspond closely to
the peaks in $V$. The shape of the $V$ profile around these bins is
also very similar between the two frequencies, but the handedness is
opposite. The examples also help to extend the significance of our
results to timescales comparable to the pulsar component widths.

\section{Summary}

In this paper we have demonstrated a method for decomposing single
pulses into the OPMs, assuming incoherently superposed OPM
emission. We support this method in light of our findings for
PSR B1133+16, where the handedness of the circular polarization has
been observed not to be perfectly associated to the dominant OPM. We
have shown that the correlation of $V$ with OPM is worse at 4.85 GHz
than at 1.41 GHz, which is consistent with the conclusion of Paper II
that propagation through the pulsar magnetosphere causes more severe
effects at higher rather than lower frequencies. Finally, we show two
examples of single pulses to demonstrate that our bin-by-bin analysis
is also applicable to structures with timescales similar to pulsar
components.

\acknowledgements{
The authors would like to thank everybody at the Effelsberg and Lovell
radio telescopes who helped in any way with the observations. AK would
like to thank the Deutscher Akademischer Austausch Dienst for their
financial support.}
\bibliographystyle{aa}
\bibliography{journals,modrefs,psrrefs}
\end{document}